\documentclass[useAMS,usenatbib]{mn2e}

\include{epsf}

\def\gsim{\lower.73ex\hbox{$\sim$}\llap{\raise.4ex\hbox{$>$}}$\,$}
\def\lsim{\lower.73ex\hbox{$\sim$}\llap{\raise.4ex\hbox{$<$}}$\,$}
\def\mpc{$\,h^{-1}\,$Mpc}
\def\%{~per~cent}

\title[2MASS Constraints on the Local Large-Scale Structure]
{2MASS Constraints on the Local Large-Scale Structure: A Challenge to $\Lambda$CDM?}
\author[W.J. Frith, T. Shanks \& P.J. Outram]
{W.J. Frith\thanks{E-mail:w.j.frith@durham.ac.uk}, 
T. Shanks \& P.J. Outram\\
Dept. of Physics, Univ. of Durham, South Road, Durham DH1 3LE, UK}
\begin{document}

\date{Accepted 2004. Received 2004; in original form 2004 }

\pagerange{\pageref{firstpage}--\pageref{lastpage}} \pubyear{2004}

\maketitle

\label{firstpage}

\begin{abstract}
We investigate the large-scale structure of the local galaxy distribution
using the recently completed 2 Micron All Sky Survey (2MASS) via three techniques. First, we
determine the $K_s$-band number counts over the $\approx$4000~deg$^2$ APM
survey area where evidence for a large-scale `local hole' has previously
been detected and compare them to a homogeneous prediction. Considering a
$\Lambda$CDM form for the 2-point angular correlation function, the
observed deficiency represents a 5$\sigma$ fluctuation in the galaxy
distribution. We check the model normalisation using faint $K$-band data compiled from the literature; the
normalisation used in this paper is in excellent agreement, and the observed counts over the APM survey area would require the model to
be lowered by 3.8$\sigma$. However, the issue is complicated by the $b\ge$20$^{\circ}$ and $b\le$-20$^{\circ}$ 2MASS
counts which lie below the best-fit model normalisation. Second, since the $K_s$-band 
counts over the APM survey area continue to suggest the possible presence
of excess clustering over the $\Lambda$CDM prediction, we next probe the
power at large scales by comparing the 2MASS and $\Lambda$CDM mock galaxy
angular power spectra. We find a $3\sigma$ excess in the 2MASS catalogue
over the $\Lambda$CDM prediction at large scales ($l\le30$). However, this
excess is not enough to account for the low counts over the APM survey
area.  Finally, in order to probe more directly whether the $\Lambda$CDM mocks can
reproduce observed features in the galaxy distribution at large scales, 
we apply a counts in cells analysis to the 2MASS data and 
mock catalogues; on the assumption that the 2MASS catalogue at
$|b|\ge20^\circ$ is representative, we find excellent agreement between
the biased $\Lambda$CDM mocks and the 2MASS catalogue to
$\approx30^{\circ}$. The crux of the interpretation of these results
appears to be whether the 2MASS volume is yet big enough to constitute a
fair sample of the Universe. The number count models based on fainter
counts suggest that it may not be, although normalisation uncertainties
remain.  It is also the case that the 2MASS depth remains comparable to
the possible size of large-scale inhomogeneities. Analyses which assume the
2MASS average density is a fair sample, such as counts in cells and to a lesser
extent, power spectral analysis, may return results which see less
contradiction with $\Lambda$CDM than the number count analysis in
the APM survey area. Further progress on assessing the significance of the local
hole and thus the consistency of the local galaxy distribution with 
$\Lambda$CDM will require deeper all-sky number counts and
redshift surveys in both the optical and the infrared.
\end{abstract}

\begin{keywords}
surveys - galaxies: photometry - cosmology: observations - large-scale structure of the
Universe - infrared: galaxies
\end{keywords}

\section{Introduction}

The counting of galaxies as a function of apparent magnitude is one of the most powerful tools 
in observational cosmology. Not only can this simple statistic form strong constraints 
on the level of evolution at the faint end, but also on the large-scale structure and the scales to 
which the cosmological principle can be said to hold from bright magnitude counts.

The APM galaxy survey \citep{mad} produced number counts which are unexpectedly low for the solid angle 
surveyed ($\approx$4000 deg$^2$). The question then arises: To what extent are 
these steep counts due to real clustering in the galaxy distribution, and do errors in the photomotery or 
strong evolution at low redshifts contribute significantly to the observed deficiency? If the counts were 
exclusively due to an under-density in the Southern 
Galactic Cap (SGC), it would be unexpectedly large and deep for our present understanding of 
large-scale structure. However, to invoke galaxy evolution alone is also problematic as it requires an extended tail in the 
$n(z)$ which is not apparent.

The presence of a large under-density in the SGC has been confirmed by large galaxy redshift surveys. The 2dF Galaxy Redshift 
Survey \citep[2dFGRS;][]{col} has taken spectra for galaxies brightward of $b_J\approx$19.45 over a solid angle of 
$>$600 deg$^2$ in the SGC. The $n(z)$ indicates remarkable structure in the local galaxy distribution with a large 
deficit to $z\approx$0.1. Large deficiencies in the $n(z)$ are also indicated by other redshift surveys in the SGC 
\citep{rat,she,vet}.

\begin{figure}
\begin{center}
\centerline{\epsfxsize = 3.5in
\epsfbox{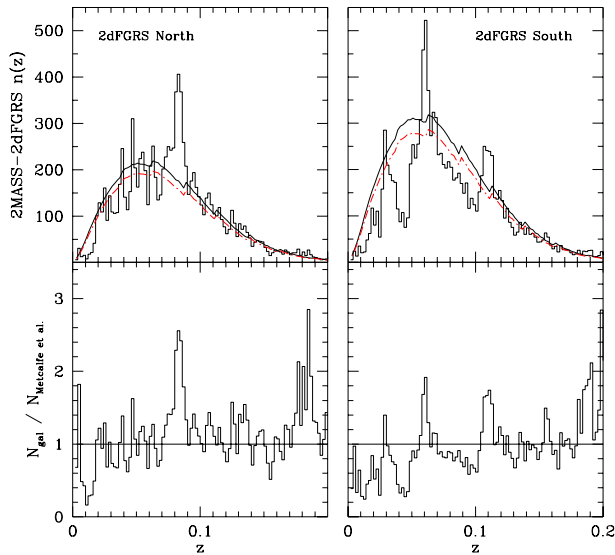}}
\caption{Number redshift histograms for the 2MASS-2dFGRS matched sample described in section 2.2 for the Northern (left-hand
panels) and Southern 2dFGRS declination strips (right-hand panels). The upper panels show the $n(z)$ for $K_s<$13.5 galaxies.
The lower panels show the corresponding radial density functions, i.e. the $n(z)$ divided through by the \citet{met2} homogeneous
prediction. In each case the \citet{met2} model is indicated by a solid line. We also show the 1$\sigma$ lower limit best fit
normalisation of the \citet{met2} $K$-band model detailed in section 3.1 (dot-dashed lines).}
\label{fig:2dfnz}
\end{center}
\end{figure}

Possible errors in the APM photometry have also been detected. Using CCD photometry, \citet{met} found a small residual 
scale error in the APM survey zero-points for $B$\gsim17. The corrected counts fainter than $B\approx$17.5 were now in 
good agreement with the expected homogeneous trend. However, a deficit in the counts of $\approx$50\% at $B\approx$16 
remained. More recently, \citet{bus} have provided $B$-band CCD photometry over $\approx$337 square degrees 
in the SGC to a limiting magnitude of $B$=20.5. The photometric zero-point is in excellent agreement 
with other CCD data such as the Millennium Galaxy Catalogue \citep[MGC;][]{dri} and the Sloan Digital Sky Survey 
\citep[SDSS;][]{yas} Early Data Release. However, a comparison with the APM Bright Galaxy Catalogue suggests a large 
zero-point offset of 0.31 magnitudes for $B<$17.35. Applying this correction to the bright APM 
photometry, \citet{bus} found that the APM survey counts now indicated a $\approx$25\% deficit at $B\approx$16, consistent with a 
25\% deficiency in the galaxy distribution to $z$=0.1 over the entire APM survey area. The implied significance using several 
possible forms for the real-space correlation function indicated that such a large local hole in the galaxy distribution 
required excess power at large scales over the $\Lambda$CDM prediction.

In conjunction with the work of \citet{bus}, \citet{fri} investigated the number counts in the $K_s$-band 
from the 2 Micron All Sky Survey (2MASS) second incremental release \citep{jar}. They found good agreement both 
with the optical number counts, and the expected trend defined by the 2dFGRS $n(z)$, suggesting that the number counts in the 
2dFGRS strips are caused by real features in the galaxy distribution, and are consistent with an absence of strong, 
low-redshift galaxy evolution. However, due to incompleteness in the 2MASS 
second incremental release, a comparison with the optical APM survey counts could not be made, although the steeper than Euclidean 
counts over the partially surveyed galactic caps indicated the possible presence of large under-densities in both the 
Southern $and$ Northern Galactic Caps to $\sim$300\mpc. This was supported by mapped 2MASS counts, showing the variations of 
counts over the sky in 5$^{\circ}\times$5$^{\circ}$ bins, and indicating huge regions of underdensity \gsim100$^{\circ}$ 
across in both hemispheres. 

Here, we probe the large-scale structure of the local galxy distribution with the recently completed 2MASS survey using 
three techniques. First, we compare the $K_s$-band 2MASS counts over the APM survey area with model predictions. We calculate 
the significance with respect to these models considering a $\Lambda$CDM form for the 2-point angular correlation function. 
Secondly, since the form of the local hole as suggested by \citet{bus} requires the presence of excess power at large scales 
over the $\Lambda$CDM prediction, we compute the 2MASS angular power spectrum and compare with a $\Lambda$CDM prediction 
constructed from the $\Lambda$CDM Hubble Volume simulation. Finally, we investigate the local galaxy distribution more 
directly by using a counts in cells analysis on the 2MASS and $\Lambda$CDM mock catalogues. 

In section 2, we present details of the datasets 
used. In section 3, the number counts are presented. The clustering at large scales is investigated through a determination 
of the 2MASS and $\Lambda$CDM mock angular power spectra in section 4. In section 5 we present the counts in cells analysis.
The discussion and conclusions follow in section 6.

\section{Data}

\subsection{The 2MASS Extended Source Catalogue}

The 2 Micron All Sky Survey (2MASS) extended source catalogue has now released $K_s$, $H$ and $J$-band photometry for over
1.6$\times$10$^6$ extended sources over the entire sky with high completeness to $K_s$=13.5 \citep{jar}. Since the second
incremental release, much of the photometry has been revised (Jarrett - priv. comm.) and the default aperture
magnitudes used in \citet{cole} have been abandoned. \citet{cole} compared the second incremental release $K_s$-band
photometry with the $K$-band photometry of \citet{lov}. The Loveday photometry had better signal-to-noise and resolution  
than the 2MASS scans and so enabled more accurate 2MASS magnitudes to be determined.
Here, we use a similar magnitude estimator to that of \citet{cole}. We take the $J$-band extrapolated magnitudes, colour-
corrected to the $K_s$-band using the $J$ and $K_s$-band fiducial elliptical magnitudes; this magnitude estimator provides the 
best fit to the \citet{lov} zeropoint as described in \citet{fri2}. An extinction correction is applied using the \citet{shl} dust maps.

\subsection{The 2dF Galaxy Redshift Survey}

The 2dF Galaxy Redshift Survey (2dFGRS) is selected in the photographic $b_J$-band using the APM survey and subsequent alterations 
and extensions to it \citep{col} for two declination strips in the northern and southern galactic caps, as well as 99 randomly 
selected 2$^{\circ}$ fields scattered over the APM survey area. The final release data has provided spectra for 
$\approx$220,000 galaxies (for a quality flag of Q$\ge$3) over $\approx$1500 square degrees to a mean magnitude limit of 
$b_J$=19.45.

In this paper, we investigate the galaxy distribution in the $K_s$-band. We have therefore formed a 2MASS-2dFGRS 
catalogue, matched over the Northern and Southern 2dFGRS declination strips, using the $K_s$-band magnitude estimator described 
above. Fig.~\ref{fig:2dfnz} shows the $K_s$-band selected $n(z)$ and radial density functions, i.e. the observed $n(z)$ over the 
predicted, for the 2dFGRS declination strips with an applied limiting magnitude of $K_s$=13.5. 

\subsection{The $\Lambda$CDM Hubble Volume Simulation}

The Hubble Volume catalogues represent the largest N-body simulations of the Universe to date. The $\Lambda$CDM simulation  
follows the evolution of 10$^9$ dark matter particles from $z\approx$50 over a volume of 3000$^3~h^{-3}$Mpc$^3$. The associated 
cosmological parameters are $\Omega_m$=0.3, $\Omega_b$=0.04, $h$=0.7, $\sigma_8$=0.9 \citep{jen}. 

In this work, we construct mock 2MASS catalogues from the $z=0$ $\Lambda$CDM Hubble Volume simulation. We divide the total volume 
into 27 virtually independent spherical volumes of $r=500$\mpc . These are subjected to the 2MASS selection function:

\begin{equation}
n(z)=\frac{3z^2}{2(\bar{z}/1.412)^3} exp \left(-\left(\frac{1.412z}{\bar{z}}\right)^{3/2}\right)
\label{equation:sel}
\end{equation}

\noindent \citep{bau,mal} where $\bar{z}$ is determined from the 2MASS-2dFGRS matched sample described in section 2.2. 
Equation~\ref{equation:sel} is normalised to match the total number of observed 2MASS galaxies for 
$|b|\ge$20$^{\circ}$. Due to the volume of the 27 mock 2MASS catalogues, the selection function is artificially truncated for 
the $K_s<$13.5 mocks at $z\approx$0.156. However, this has a negligible effect on the work in this paper; at this redshift, 
$\approx$95\% of the galaxies are sampled for $K_s<$13.5.

For the counts in cells analysis, it is necessary to more accurately mimic the galaxy sample. We therefore use a bias 
prescription:

\begin{equation}
P(\nu) =  \left\{ \begin{array}{ll}
   \exp(\alpha\nu + \beta\nu^{3/2}) & \mbox{for $\nu\geq0$} \\ 
    \exp(\alpha\nu) & \mbox{for $\nu<0$},
     \end{array} \right. 
\label{equation:bias}
\end{equation}

\noindent \citep{cole2} where $P(\nu)$ is a bias probability based on the density field and $\nu$ is the number of standard 
deviations of the density away from the mean. For the counts in cells analysis we use magnitude-limited data of $K_s<$12.5. We 
use parameters of $\alpha$=0.6 and $\beta$=-0.15 to match the $K_s<$12.5 angular correlation function at small scales 
($\theta$\lsim10$^{\circ}$).

\section {Number Counts}

\subsection{Model Normalisation}

The issue of the local hole rests critically on the model number count normalisation. In the $K$-band, the number count 
predictions for passively-evolving galaxies are in remarkable agreement with the observations to $K\approx$23 \citep{mcc}. The 
predicted number below $K\approx$18 is also fairly insensitive to the evolutionary model or the assumed cosmology. Therefore, the 
$K$-band number counts are a particularly useful probe of the local Universe since the model predictions can be constrained at 
fainter magnitudes with few concerns over uncertainties in the amount of evolution or the cosmology. 

In this work, we use a non-evolving $K$-band model computed from the luminosity function parameters of \citet{met2} and the 
$K$-corrections of \citet{bru}. Comparing the number count predictions in the fitting range 14$<K<$18 to the faint counts used 
in \citet{mcc} as well as other data \citep{vai,szo,hua,hua2,kum,mar,mcl}, we find that the \citet{met2} model provides a 
good fit to the observations with $Y$=0.96$\pm$0.06 (where $Y$=1 represents the \citet{met2} normalisation). 

In order to test for the presence of strong evolution at low redshifts and zero-point offsets in the 2MASS data, we also 
construct model predictions for the number counts from the \citet{met2} homogeneous prediction described above and the observed 
$n(z)$ \citep{fri}. These are determined by varying the luminosity function normalisation as a function of the redshift; the 
luminosity function parameter $\phi^*$ is multiplied by the relative density (Fig.~\ref{fig:2dfnz}, lower panels). These variable 
$\phi^*$ models therefore provide a simple prediction for the number counts for a given galaxy distribution, assuming that there 
is no significant effect from strong galaxy evolution at low redshifts.

\subsection{Results}

\begin{figure}
\begin{center}
\centerline{\epsfxsize = 3.5in
\epsfbox{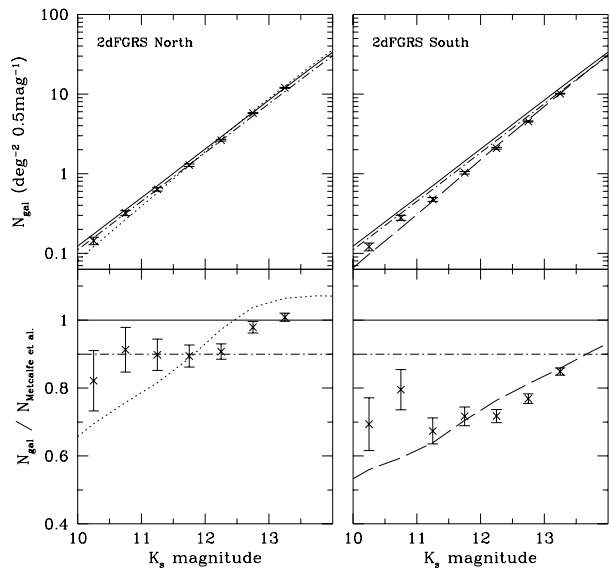}}
\caption{$K_s$-band 2MASS number counts extracted for the 2dFGRS Northern (left-hand panels) and Southern (right-hand panels)
declination strips. The upper panels show the counts on an ordinary logarithmic plot, while the lower panels show the residual
plot of the counts, i.e. the number count divided through by the \citet{met2} homogeneous prediction. The homogeneous model is
indicated by a solid line, with the 2dFGRS Northern and Southern variable $\phi^*$ models shown by the dotted and dashed lines
respectively. We also show the 1$\sigma$ lower limit best fit normalisation of the \citet{met2} $K$-band
model detailed in section 3.1 (dot-dashed lines). The errorbars are Poissonian.}
\label{fig:2dfnm}
\end{center}
\end{figure}

In order to verify the consistency of bright number counts with the corresponding $n(z)$ \citep{fri}, we compare the 
$K_s$-band number counts extracted from the 2dFGRS strips with the corresponding variable $\phi^*$ models, using the 
newly-completed 2dFGRS and 2MASS catalogues (Fig.~\ref{fig:2dfnm}). 
The $K_s$-band number counts extracted from the 2dFGRS fields are in reasonable agreement with the corresponding 
variable $\phi^*$ models. This suggests that the form of the bright number counts is exclusively due to features in the local 
galaxy distribution and is consistent with an absence of strong galaxy evolution at low redshifts. The agreement between the 
counts and the variable $\phi^*$ models is independent of the model normalisation, since any change in the number count models
also alters the $n(z)$ model normalisation and therefore the implied deficiency to the same degree.

\begin{figure}
\begin{center}
\centerline{\epsfxsize = 3.5in
\epsfbox{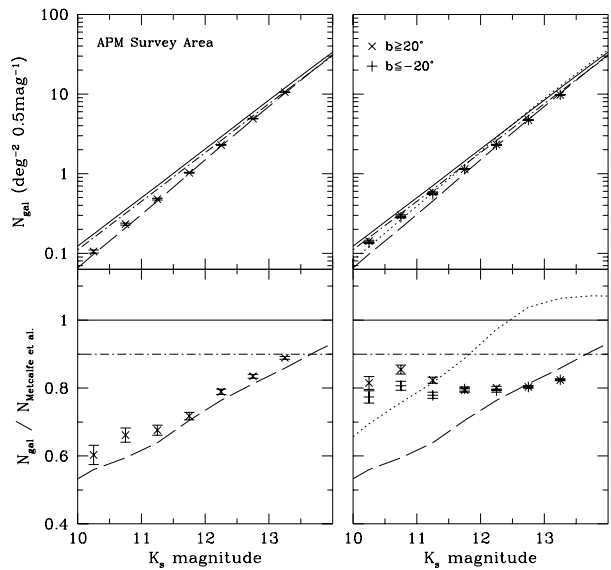}}
\caption{$K_s$-band 2MASS number counts extracted for the $\approx 4000$ deg$^2$ APM survey area and the $|b|\ge 20^{\circ}$
galactic caps, presented as in Fig.~\ref{fig:2dfnm}}.
\label{fig:apm}
\end{center}
\end{figure}

Having confirmed the consistency between the number counts and the underlying large-scale structure in the 2dFGRS strips, we 
are now in a position to examine the number counts over the APM survey area. Fig.~\ref{fig:apm} shows the 
$K_s$-band 2MASS counts extracted for the $\approx$4000 deg$^2$ field. We also show the 2dFGRS Southern variable 
$\phi^*$ model for reference. 

There is a surprisingly good agreement between the $K_s$-band counts and the 2dFGRS Southern variable $\phi^*$ model. This 
suggests that the local galaxy distribution in the APM survey area may be similar to that of the much smaller 2dFGRS Southern 
strip. This is supported by similar deficiencies 
observed in the Durham-UKST redshift survey \citep{rat}, the Las Campanas Redshift Survey \citep{she} and the ESO Slice Project 
\citep{vet}, which are all situated within the APM survey area, and also the optical number counts and corresponding 
variable $\phi^*$ models \citep{fri3, fri, bus}. The degree of under-density may 
therefore be inferred from the observed 2MASS-2dFGRS redshift distribution and the model normalisation. Taking the \citet{met2} 
$K$-band model normalisation and the 2MASS-2dFGRS matched $n(z)$ (Fig.~\ref{fig:2dfnz}) implies a 24\% under-density to $z$=0.1. 
If we take the $1\sigma$ lower limit for the best fit $K$-band model normalisation detailed in section 3.1, the implied 
under-density over the APM survey area to $z$=0.1 is 15\%.

In order to see whether this structure persists to even larger scales than the APM survey area, we have determined the 2MASS 
$K_s$-band counts over the entire $|b|\ge$20$^{\circ}$ galactic caps (Fig.~\ref{fig:apm}). The counts are very low with respect 
to the model normalisations used previously, but are in good agreement with each other. This may indicate that the normalisation 
of the \citet{met2} model should be even lower, 2.5$\sigma$ below the best fit to the $K$-band counts at 14$<K<$18. 
Alternatively, it might also indicate the presence of a zero-point offset between the 2MASS photometry and the $K$-band model of 
$\approx$0.15 magnitudes; any alteration to the 2MASS zero-point would compromise the agreement between the $K_s$-band 2dFGRS strip 
counts and the corresponding variable $\phi^*$ models, and also with the \citet{lov} photometry zero-point with which the 2MASS 
magnitudes are compared. If the low counts over the galactic caps were due to real features in the galaxy distribution, this 
would imply that the local Universe is globally under-dense, and that even surveying over $\approx$25,000 deg$^2$ to 
$r$\gsim150\mpc\  does not constitute a fair sample of the Universe.

 
\subsection {Determining the Significance}

\begin{table}
\centering
\begin{tabular}{||c|c|c|c||} \hline
Model Normalisation   &	$Y$	& Deficiency             & Significance \\
\hline
Metcalfe et al.       &	1.00	& 26.9\%                 & 5.0$\sigma$   \\
Best fit              &	0.96	& 23.9\%                 & 4.5$\sigma$   \\
1$\sigma$ lower limit &	0.90	& 18.8\%                 & 3.5$\sigma$   \\
$\bar{n}_{~2MASS}$    &	0.80	& 9.7\%                  & 1.8$\sigma$   \\
\hline

\end{tabular}
\caption{\small{Significances for the observed deficiency in the $K_s$-band counts extracted for the APM survey area
for $K_s<$12.5 considering various model normalisations (the relative normalisations are indicated by $Y$). In each case a $\Lambda$CDM 
correlation function is considered of A=0.28 1-$\gamma$=-0.71
for $\theta<$5.0$^{\circ}$, A=5.3, 1-$\gamma$=-2.5 for $\theta \ge$5.0$^{\circ}$ where
$\omega=A\theta^{1-\gamma}$. The best fit parameters at small scales are taken from \citet{mal}, while the large scale fit
is determined from the $K_s<$12.5 correlation function presented in Fig.~\ref{fig:2mass_corr} for $\theta\ge$10$^{\circ}$. This is
formed from the 27 mock 2MASS catalogues constructed from the $\Lambda$CDM Hubble Volume mocks with a bias of $b_K$=1.1
\citep{mal} and is in good agreement with the \citet{mal} power law best fit at $\theta \approx$1$^{\circ}$. }}
\label{table:sig}
\end{table}

Using assumed forms to the galaxy correlation function at large scales, it is possible to determine the associated 
significance of features in the galaxy distribution. \citet{bus} considered various forms to the 
real-space correlation function to calculate the significance of an assumed 3-dimensional form to the local hole. Here, we use 
the 2-dimensional analogue to determine the significance implied by the $observed$ deficiency in the number counts with respect 
to various normalisations of the \citet{met2} model using the angular correlation function:

\begin{equation}
\left(\frac{\delta N}{\bar{N}}\right)^2=\frac{1}{\bar{N}}+\frac{1}{\Omega^2}\int d\Omega_1d\Omega_2\omega (\theta_{12})
\label{equation:sig}
\end{equation}

\noindent \citep{peb4} where $\omega (\theta_{12})$ is the value of the two-point angular correlation function between two area 
elements $d\Omega_1$ and $d\Omega_2$. $\Omega$ is the total solid angle of the survey, $n$ is the mean galaxy density such that 
$\bar{N}=n\Omega$ is the total number of galaxies in the survey area. A power law form of the correlation function is assumed 
such that $\omega=A\theta^{1-\gamma}$.

In Table~\ref{table:sig}, we apply this technique to the observed deficiency over the APM survey area in the 2MASS $K_s$-band 
counts for $K_s<$12.5 with respect to the \citet{met2} model, the best fit and the 1$\sigma$ lower limit best fit normalisations of 
this model described in section 3.1, and the observed mean 2MASS number count for $|b|\ge20^{\circ}$ (see Fig.~\ref{fig:apm}). 
This magnitude limit is chosen such that the 
peak in the selection function lies within the redshift range of interest; at $K_s<$12.5 the $n(z)$ peaks at $z\approx$0.05, and 
so this magnitude limit preferentially samples the mean depth of the possible local hole in the APM survey area. We 
consider a $\Lambda$CDM form for the angular correlation function determined from fits to the mock 2MASS    
catalogues (see section 4.3).

Both the \citet{met2} model and the best fit normalisation appear to imply that the observed 2MASS 
number counts over the APM survey area represent either an extremely rare fluctuation in the galaxy density or a challenge to the 
large-scale clustering predicted by the $\Lambda$CDM Hubble Volume simulation. Even lowering the normalisation to the 1$\sigma$ 
best fit lower limit yields a significance of $>3\sigma$. Only if the \citet{met2} model is lowered by $2.5\sigma$ to the mean 
2MASS number density at $K_s<$12.5 for $|b|\ge20^{\circ}$ ($\bar{n}_{~2MASS}$) , do the counts 
become more consistent with the $\Lambda$CDM prediction. To remove any discrepancy in the $K_s<$12.5 counts over the APM survey area 
requires the model normalisation to be lowered by 3.8$\sigma$. Combining the error on the normalisation and the significance estimate 
for the best fit normalisation in quadrature yields a significance of 3.0$\sigma$. Clearly, unless the faint $K_s$-band counts are 
systematically too high, then the 2MASS counts over the APM survey area suggest the presence of excess power at large scales over the 
$\Lambda$CDM prediction.

\section {The 2MASS Angular Power Spectrum}

The large local hole in the APM survey area, as suggested by \citet{bus} (a $\approx$25\% deficiency to $z$=0.1 over 
$\approx$4000 deg$^2$) and the $K_s<$12.5 2MASS number counts, appear to imply the presence of excess power at 
large scales over the $\Lambda$CDM prediction. In order to investigate this possible presence of excess power at large 
scales, we compute the angular power spectrum for $|b|\ge$20$^{\circ}$ 2MASS galaxies and 2MASS N-body mocks constructed 
from the $\Lambda$CDM Hubble Volume mock catalogue. An examination of the implied cosmological parameters and of various possible sources 
of systematic error as well as a more detailed description of the method can be found in \citet{fri2}.

\subsection{Method}

Following the usual method \citep[e.g.][]{peb,peb2,peb3,sch}, the angular 
power is estimated through a spherical harmonic expansion of the surface density of galaxies. The coefficients of this 
expansion are determined over the observed solid angle $\Omega_{obs}$:

\begin{equation}
a_l^m =\sum_{N_{gal}} Y_l^m(\theta ,\phi ) - {\mathcal{N}} \int_{\Omega_{obs}} Y_l^m(\theta ,\phi ) d\Omega
\end{equation}

\noindent where ${\mathcal{N}}$=$N_{gal}/\Omega_{obs}$ is the observed number of galaxies per steradian. The angular power
is then determined as the ratio of the observed signal to the expected Poisson fluctuation, ${\mathcal{N}}$:

\begin{equation}
C_l = \frac{1}{{\mathcal{N}}(\Omega_{obs}/4\pi)(2l+1)} \sum_m |a_l^m|^2
\label{equation:norm}
\end{equation}

\noindent such that $C_l$=1 corresponds to a random distribution for all-sky coverage. This is described in more detail in 
\citet{fri2}.

\subsection{Results}

\begin{figure}
\begin{center}
\centerline{\epsfxsize = 3.5in
\epsfbox{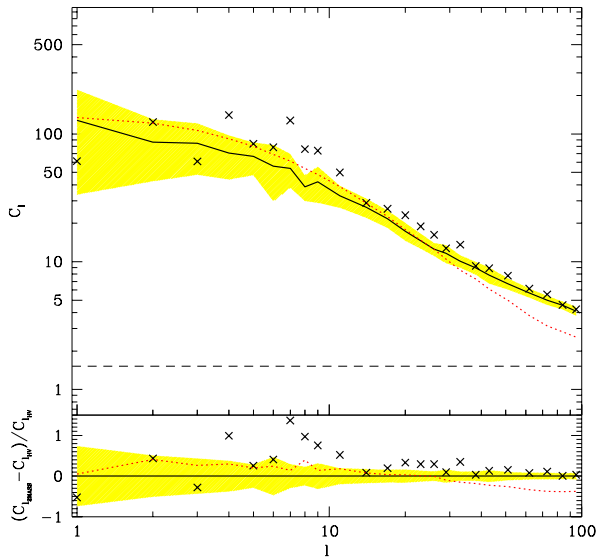}}
\caption{The $|b|\ge$20$^{\circ}$ 2MASS angular power spectrum for 124,264 $K_s<$12.5 galaxies. The crosses indicate
the 2MASS datapoints with the shaded region and solid line indicating the 1$\sigma$ spread and mean power spectrum of 27 mock
2MASS catalogues constructed from the $\Lambda$CDM Hubble Volume mock catalogue. The model corresponding to the Hubble
Volume mock catalogue input parameters of $\Omega_m$=0.3, $\Omega_b$=0.04, $h$=0.7 and $\sigma_8$=0.9 is indicated by the
dotted line. The upper panel shows the power spectrum normalised as in equation~\ref{equation:norm} with the dashed line
indicating the expected Poisson fluctuation for the solid angle used. In the lower panel we show the 2MASS, mock and
model power spectra expressed as the fractional deviation from the mean mock power spectrum.}
\label{fig:2mass_mag}
\end{center}
\end{figure}

\begin{figure}
\begin{center}
\centerline{\epsfxsize = 3.5in
\epsfbox{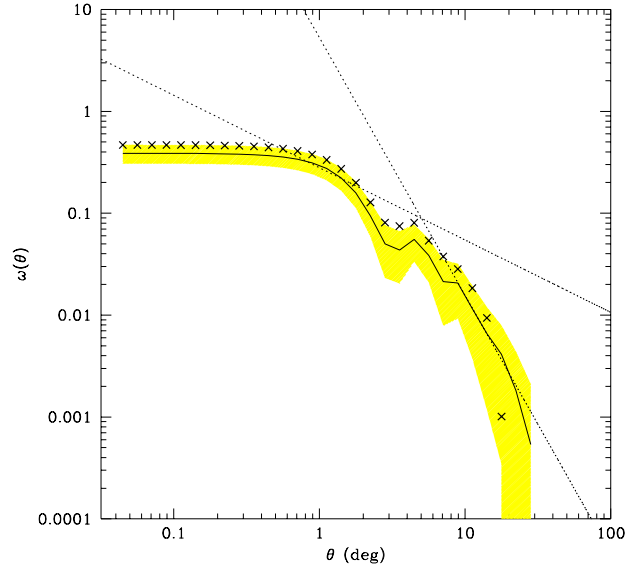}}
\caption{The $K_s<$12.5, $|b|\ge$20$^{\circ}$ 2MASS (crosses) and mock 2MASS (solid line) 2-point angular correlation functions determined
via a Bessel function transform of the $K_s<$12.5 angular power spectrum for 1$\le l\le$100. We have included a bias of $b_K$=1.1
for the mock 2MASS catalogues. The solid lines indicate power law fits of A=0.28, 1-$\gamma$=-0.71 for $\theta<$4.5$^{\circ}$,
A=4.4, 1-$\gamma$=-2.5 for $\theta \ge$4.5$^{\circ}$ where $\omega=A\theta^{1-\gamma}$. The small scale fit is 
taken directly from \citet{mal}; at large scales the fit is determined for $\theta\ge$10$^{\circ}$.}
\label{fig:2mass_corr}
\end{center}
\end{figure}

The 2MASS angular power spectrum for $\approx$120,000 $K_s<12.5$, $|b|\ge$20$^{\circ}$ galaxies 
is presented in Fig.~\ref{fig:2mass_mag}. We also show the mean and 1$\sigma$ spread determined from 27 mock 2MASS catalogues 
described in section 2.4. The linear model corresponding to the $\Lambda$CDM Hubble Volume mock catalogue input 
parameters, corrected for the window, is indicated by the dotted line.

The 2MASS angular power spectrum is in reasonable agreement with the mock 2MASS angular power spectra although the 2MASS angular 
power spectrum slope is steeper and there is some discrepancy with the mock catalogue in an unbiased scenario. 
Therefore it appears that either there is an excess of power in the 2MASS catalogue over the $\Lambda$CDM 
Hubble Volume or there exists a scale-dependent bias within the scales shown. Clearly, the issue of bias is critical in 
determining the level of disagreement at large scales. Taking a reasonable value of the $K_s$-band bias of $b_K$=1.1 
\citep{mal}, the disagreement at large scales ($l\le$30; this corresponds to $r$\gsim30\mpc\  at the mean depth of the 
$K_s<$12.5 sample) is at the $\approx$3$\sigma$ level.

\subsection{The relevance for a large local hole}

Using equation~\ref{equation:sig}, we can determine whether the possible excess of power observed at large scales in the 2MASS 
angular power spectrum over the $\Lambda$CDM prediction can account for a large local hole in the APM survey area, via a 
transform of the angular power spectrum to the angular correlation function. For this we use a Bessel function transform 
\citep{efs}:

\begin{equation}
\omega(\theta )\approx \frac{1}{2\pi } \sum_l ~l ~C_l ~J_0(l\theta )
\end{equation}

In Fig.~\ref{fig:2mass_corr}, we present the transformation of the $K_s<12.5$ 2MASS and $\Lambda$CDM mock 
2MASS power spectra shown in Fig.~\ref{fig:2mass_mag} to the angular correlation function, together with 
the best fit power laws used in section 3.3. Both the 2MASS and mock 2MASS angular correlation functions are in good 
agreement with the \citet{mal} best fit slope at $\approx1^{\circ}$; at smaller scales, the transformation becomes unreliable due 
to the lack of angular power spectrum information beyond $l$=100.

The 2MASS angular correlation function is in good agreement with the mock 2MASS catalogues at large scales, although there is a 
small difference in slope. Computing the significance as in Equation~\ref{equation:sig} using a new best fit power law to the 
2MASS angular correlation at large scales ($\theta\ge10^{\circ}$) does slightly reduce the significance; the significance 
estimates in Table~\ref{table:sig} are reduced by \lsim0.5$\sigma$ in each case. 

\begin{figure*}
\begin{center}
\centerline{\epsfxsize = 7in
\epsfbox{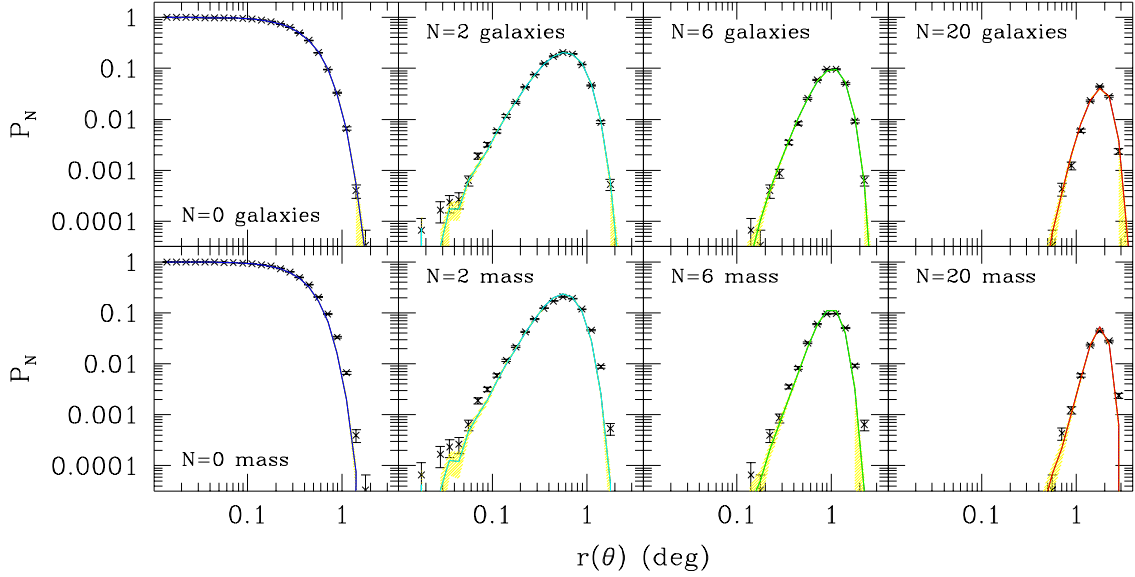}}
\caption{Count Probability Distribution Functions (CPDF) for $N$=0, 2, 6 and 20 for $K_s<$12.5 2MASS
galaxies (crosses). The mean CPDFs (solid line) and 1$\sigma$ spread from the 27 mock 2MASS catalogues are also shown, for
unbiased (lower panels) and biased (upper panels) particles. The errorbars are Poissonian.}
\label{fig:cpdf1}
\end{center}
\end{figure*}

\begin{figure*}
\begin{center}
\centerline{\epsfxsize = 7in
\epsfbox{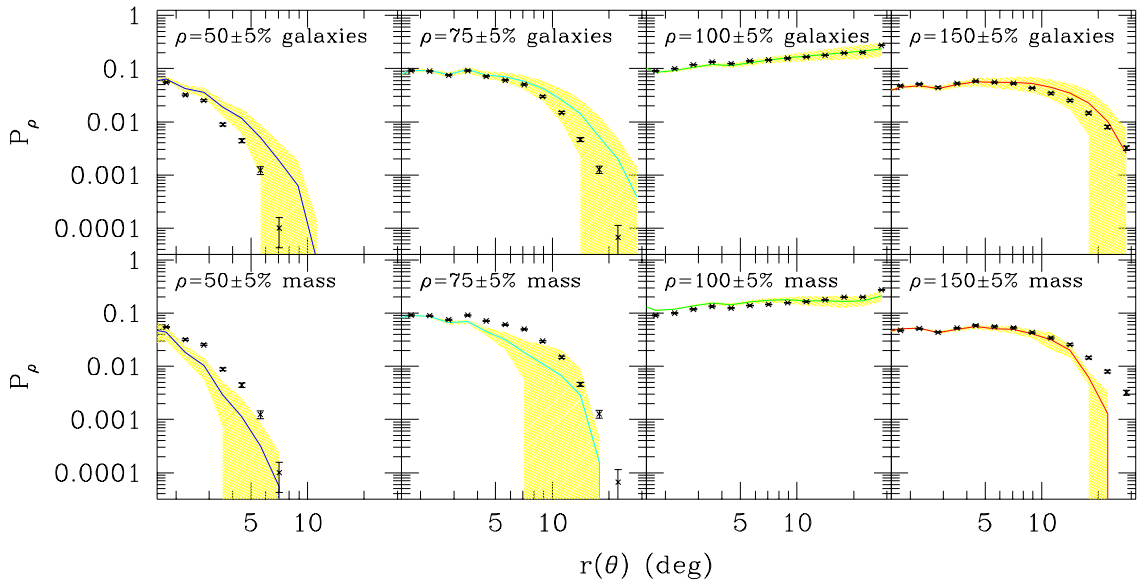}}
\caption{Density Probability Distribution Functions (DPDF) for densities of $\rho$=50$\pm$5\%, 75$\pm$5\%, 100$\pm$5\% and
150$\pm$5\% for $K_s<$12.5 2MASS galaxies (crosses). The mean DPDFs (solid line) and 1$\sigma$ spread from
the 27 mock 2MASS catalogues are also shown, for unbiased (lower panels) and biased (upper panels) particles. The errorbars are
Poissonian.}
\label{fig:cpdf2}
\end{center}
\end{figure*}

Since the significance estimates detailed in Table~\ref{table:sig} are only marginally effected by this change in the large-scale slope 
of the angular correlation function, it appears that the discrepancy found between the 2MASS angular power spectrum and the $\Lambda$CDM 
mock 2MASS power spectra at large scales is not able to account for the large local hole in the galaxy distribution described in section 
3. The only way in which this local hole and the lack of excess large-scale power over the $\Lambda$CDM prediction can be reconciled is 
if the whole local sample is significantly biased by the local galaxy distribution in the APM survey area and therefore that the 2MASS 
catalogue is not a fair sample of the Universe. This is further discussed in section 6.

\section{Counts in Cells}

Our third technique to investigate the local large-scale structure is a counts in cells analysis of the $K_s<12.5$ 2MASS sample. At 
one level this is simply a check of the significance calculation used in section 3.3. However, it is also interesting to 
investigate how higher order moments in the galaxy clustering may effect the large-scale structure, and also whether the Hubble 
Volume simulation is able to reproduce the observed structure over cosmologically significant volumes.

\subsection{Method}

We sample the 2MASS survey area with a large number of randomly placed circular cells with an angular radius $r$. The Count 
Probability Distribution Function (CPDF) is defined as the probability of finding an exact number of galaxies, $N$, in a 
particular cell as a function of the cell size \citep[e.g.][]{cro}. Here, we determine the CPDFs for $N$=0, 2, 6 and 20 sampling 
scales of \lsim3$^{\circ}$. We therefore limit the area over which the cells are placed to $|b|\ge$20$^{\circ}$ and the 2MASS 
$K_s<$12.5 sample to $|b|\ge$10$^{\circ}$ to remove spurious edge effects. This magnitude limit is chosen in order that the peak 
in the selection function lies within the redshift range of interest; at $K_s<$12.5 the $n(z)$ peaks at   
$z\approx$0.05, and so this magnitude limit preferentially samples the mean depth of the possible local hole in the APM survey
area.

In order to probe much larger scales, of interest in this work, it is more useful to probe the density of galaxies rather than 
the absolute number. We define a Density Probability Distribution Function (DPDF) as the probability of finding a cell of given 
density (determined with respect to the mean 2MASS density) as a function of cell size. Since we wish to probe the galaxy 
distribution at large angular scales, we limit the cells to $|b|\ge$42$^{\circ}$ and the 2MASS $K_s<$12.5 sample to 
$|b|\ge$10$^{\circ}$ as before. For both the CPDFs and DPDFs, we mimic the 2MASS sample with the 27 biased and unbiased mock 
2MASS catalogues described in section 2.3.

\subsection{Results}

In Fig.~\ref{fig:cpdf1}, we show the CPDFs for $N$=0, 2, 6 and 20 determined for the 2MASS and mock 2MASS samples. The agreement 
between the 2MASS and the mock galaxy samples is excellent, although this is not unexpected since the CPDF is dominated by the 
2-point correlation function; the mock galaxy sample is produced to match the observed $\omega(\theta)$ at small scales.

Moving to larger scales, Fig.~\ref{fig:cpdf2} shows the DPDFs for the 2MASS and mock galaxy samples for a range of densities, 
determined with respect to the mean 2MASS density. Again, the agreement is excellent to extremely large scales 
($\theta$\lsim30$^{\circ}$). It is also interesting to note that the introduction of bias has an extremely significant effect on 
the resulting DPDF. While the solid angles probed in the largest bin are slightly smaller than the APM survey area, the good 
agreement between the biased mocks and the 2MASS sample confirms the significance calculation shown in Table~\ref{table:sig} 
which suggests that, when compared to the mean $|b|\ge$20$^{\circ}$ 2MASS number count, the observed deficiency in the APM survey 
area is not significant when compared to the $\Lambda$CDM prediction. The mock 2MASS catalogues are normalised to the mean number 
count observed by 2MASS and so any increase in the global galaxy density, as might be suggested by the faint $K$-band counts, increases 
the significance of a large local hole. 

\section {Discussion \& Conclusions}

Recent evidence \citep{bus,fri,fri3} has suggested that while the optical number counts over the APM survey area may be 
significantly less deficient than originally proposed, the resulting under-density might still present a challenge to the 
form of clustering predicted by $\Lambda$CDM at large scales. In this paper, we have presented three different methods of 
analysis to probe the possible existence of this large local hole in the galaxy distribution in the SGC using the recently 
completed 2MASS survey. 

First, we determined the $K_s$-band number counts over the APM survey area. In order to probe the underlying galaxy 
distribution, we compared the APM counts with a model guided by the Southern 2dFGRS $n(z)$; this variable $\phi^*$ model provides 
a reasonable agreement with the corresponding $K_s$-band counts extracted for the Southern 2dFGRS strip. The agreement between 
the Southern 2dFGRS variable $\phi^*$ model and the $K_s$-band counts over the APM survey area is remarkable, and suggests that 
the galaxy distribution over this $\approx$4000 deg$^2$ area may be similar to that of the much smaller 2dFGRS Southern strip. 
Using the 2MASS-2dFGRS matched $n(z)$, this would imply a mean deficiency in the galaxy distribution of 24\% to $z=0.1$ with 
respect to the \citet{met2} model used in this paper; taking the 1$\sigma$ lower limit best fit normalisation of this model to 
faint $K$-band data compiled from the literature in the range 14$<K<$18 implies a mean deficiency of 15\% to $z$=0.1.

The issue is complicated by the $b\ge$20$^{\circ}$ and $b\le$-20$^{\circ}$ 2MASS counts which are in good agreement with each 
other but are significantly below the \citet{met2} homogeneous prediction. There are three possible interpretations. The first 
is that the model normalisation is too high. However, if the model were scaled down in order to agree with the mean 2MASS 
number density, this would require a 2.5$\sigma$ deviation from the faint $K$-band counts; to account for the low APM counts 
entirely through a change in the normalisation would require a 3.8$\sigma$ deviation. Secondly, the low cap counts might 
indicate a zero-point offset between the \citet{met2} model and the 2MASS data. Invoking the $\approx$0.15 magnitude shift necessary 
to bring the cap counts and the model into line would compromise the good agreement between the 2MASS and \citet{lov} photometry 
and also between the 2dFGRS $K_s$-band counts and the corresponding variable $\phi^*$ models. Thirdly, the low 2MASS cap counts 
might indicate that the entire local galaxy distribution is globally under-dense. While this might appear to be unlikely, the 
observed 2MASS counts over the APM survey area suggest that large inhomogeneities exist in the galaxy distribution over 
extremely large volumes, and so perhaps only a few such features are necessary to bias the entire local sample. It is therefore 
not possible to rule out this final possibility without further analysis. However, since the $b\ge$20$^{\circ}$ and 
$b\le$-20$^{\circ}$ counts are similar, this position requires the coincidence that we are positioned in the centre of this local 
underdensity \citep{lov2}.

In order to determine the significance of the observed $K_s$-band counts over the APM survey area, we calculated the expected 
variance over $\approx$4000 deg$^2$ considering a $\Lambda$CDM form of the 2-point angular correlation function at large scales. 
The observed deficiency is calculated with respect to the \citet{met2} model, the best fit and 1$\sigma$ lower limit best fit 
normalisations of this model described in section 3.1, and the mean 2MASS $|b|\ge20^{\circ}$ number density ($\bar{n}_{~2MASS}$). In 
the first three cases, the observed counts represent an extremely rare fluctuation from that expected by $\Lambda$CDM. If the 
\citet{met2} model is effectively lowered by 2.5$\sigma$ with respect to the faint $K$-band data to the mean 2MASS number density, 
then the observed counts begin to become more consistent with a $\Lambda$CDM form of the correlation function.
Therefore, unless the faint $K$-band data are systematically too high or the galaxy distribution in the SGC 
is an extremely rare fluctuation in the galaxy density, then the $K_s$-band counts over the APM survey area appear to imply an 
excess of power at large scales over the $\Lambda$CDM prediction. 

Our second technique is therefore to examine the large-scale power in the 
2MASS catalogue through a determination of the galaxy angular power spectrum. We compare this to a $\Lambda$CDM prediction 
determined from the Hubble Volume simulations. The two are in reasonable agreement although there is some discrepancy in the 
slopes; taking a $K_s$-band bias of $b_K$=1.1 \citep{mal} results in a $\approx3\sigma$ excess over the mean $\Lambda$CDM angular 
power spectrum at large scales ($l\le$30). In order to determine the effect of this apparent excess on the significance estimates 
used previously, we transform the angular power spectrum to the angular correlation function via a Bessel function transform. The 
corresponding best fit to the 2MASS angular correlation function at large scales ($\theta\ge10^{\circ}$) decreases the 
significance estimates by \lsim0.5$\sigma$ compared to the $\Lambda$CDM angular correlation function used previously. Therefore 
while there appears to be an excess of power at large scales in the 2MASS catalogue over the $\Lambda$CDM Hubble Volume simulation, 
it is not enough to account for the observed deficiency in the APM survey area. One caveat to this is that the mock 2MASS 
catalogues are constructed such that the mean galaxy density agrees with that of the $|b|\ge20^{\circ}$ 2MASS sample. If, as might 
be inferred from the low cap counts in Fig.~\ref{fig:apm}, the local galaxy distribution was globally under-dense with respect to 
the faint $K$-band counts, then the corresponding significance estimates would change due to the fact that the 2MASS correlation 
function is not drawn from a fair sample of the Universe.

Our third technique is to use a 2-dimensional counts in cells analysis on the 2MASS catalogue and also the unbiased and biased 
mock 2MASS catalogues constructed from the $\Lambda$CDM Hubble Volume simulation (again normalised to the mean 2MASS galaxy 
density). This enables us to verify the significance estimates used previously, and also that the form of real features in the 
galaxy distribution are reproduced by the $\Lambda$CDM Hubble Volume simulation at large scales. The biased mock catalogues 
reproduce the observed galaxy distribution to very large scales (\lsim30$^{\circ}$). This supports the significance calculation 
determined previously (with respect to the mean 2MASS number density) which indicates that the local hole is not a challenge to 
$\Lambda$CDM if the $K_s<12.5$ 2MASS catalogue is a fair sample of the Universe.

In conclusion, the issue of the large local hole in the local galaxy distribution has yet to be resolved. The 2MASS $K_s$-band 
number counts extracted for the $\approx$4000 deg$^2$ APM survey area indicate a clear deficiency in the local galaxy 
distribution consistent with the form of the $n(z)$ in the much smaller 2dFGRS Southern strip. However, to determine the level of 
the deficiency in the galaxy distribution requires an accurate normalisation of the $K$-band number count model. Using the 
\citet{met2} model, which provides an excellent fit to faint $K$-band data compiled from the literature in the range 14$<K<$18, 
implies a deficiency over the APM survey area which is at odds with $\Lambda$CDM and a local galaxy distribution which is 
globally under-dense. Only if the model is lowered by 3.8$\sigma$ below the faint $K$-band data can the normalisation account for 
the low counts over the APM survey area. The large increase in faint $K$-band data from the UK Infra-red Deep Sky Survey should 
help to resolve this issue.

\section*{Acknowledgements} 

This publication makes use of data products from the 2 Micron All-Sky Survey, which is a joint project of the University of
Massachusetts and the Infrared Processing and Analysis Centre/California Institute of Technology, funded by the Aeronautics and
Space Administration and the National Science Foundation. We thank Adrian Jenkins and Carlton Baugh for their assistance with the 
Hubble Volume mock catalogues and useful discussion. We also thank Tom Jarrett for his help with the 2MASS magnitudes.




\label{lastpage}

\end{document}